\documentclass[twocolumn]{aastex7}

\usepackage[T1]{fontenc}
\usepackage{graphicx}
\usepackage{amsmath}
\usepackage{placeins}
\usepackage{xcolor}
\definecolor{BrickRed}{rgb}{0.8,0.1,0.1}

\newcommand{\figfile}[2][]{%
  \IfFileExists{#2}{\includegraphics[#1]{#2}}{%
    \fbox{\parbox[c][2.2in][c]{\linewidth}{\centering Missing figure file:\\\texttt{#2}}}%
  }%
}

\newcommand{\Msun}{M_\odot}

\shorttitle{Rest-frame $K$-band view of bulge growth at $z\sim2$}
\shortauthors{Tadaki}

\begin{document}

\title{A Rest-frame $K$-band View of Bulge Growth in Massive Star-forming Galaxies at $z\sim2$}

\author{Ken-ichi Tadaki}
\email{tadaki@hgu.jp}
\affiliation{Faculty of Engineering, Hokkai-Gakuen University, Toyohira-ku, Sapporo 062-8605, Japan}

\begin{abstract}
We study bulge formation in 16 massive dusty star-forming galaxies at $z\sim2$ by combining JWST MIRI F770W imaging, NIRCam imaging in eight bands, and ALMA 870~$\mu$m data.
At the median redshift of $z=2.18$, the F770W band probes the rest-frame 2.4~$\mu$m light, close to the rest-frame $K$ band, and traces the stellar mass distribution with much less dust attenuation than the NIRCam bands.
The effective radii measured in F770W agree with those of the 870~$\mu$m dust emission, with a median ratio of 0.95, and are typically 23\% smaller than those in F444W.
The rest-frame $K$-band imaging is therefore essential to measure the true extent of the stellar mass, which is as compact as the dusty star-forming regions.
In stacks of the 16 galaxies at a matched resolution, the 870~$\mu$m and F770W profiles agree closely out to 8~kpc, and a bulge plus disk decomposition of the F770W stack gives a bulge-to-total ratio of 0.50.
Fitting the SEDs in radial bins with CIGALE, we find that the attenuation decreases from $A_V=3.7$~mag at the center to 1.9~mag at $r\gtrsim7$~kpc, while the specific star formation rate profile is flat.
In most of the galaxies, the bulge is already in place inside the dusty core.
In two galaxies, by contrast, the 870~$\mu$m emission is 3 to 6 times more compact than the F770W light, which suggests that they are still in the bulge-building compaction phase.
Our sample may thus catch massive galaxies in several phases of bulge formation.
\end{abstract}


\section{Introduction} \label{sec:intro}

Most massive galaxies in the local universe are dominated by a stellar bulge, and when and how these bulges formed is a central question in galaxy evolution.
Observations of massive quiescent galaxies at $z\sim2$ give an important clue.
They are extremely compact, with effective radii of 1--2~kpc \citep{Daddi2005,vanDokkum2008,vanderWel2014}, and their central stellar densities are already similar to those of the cores of local early-type galaxies \citep{vanDokkum2014,Barro2017}.
The bulges of massive galaxies must therefore be built during the star-forming phase at higher redshifts, around the peak of the cosmic star formation history \citep{MadauDickinson2014}.

Several paths have been proposed for this early bulge growth.
Gas-rich mergers can drive gas to the center and trigger a nuclear starburst \citep{Mihos1996,Hopkins2008}.
Internal processes can do the same without a merger.
In a gas-rich disk that is gravitationally unstable \citep{Romeo2010}, massive clumps form and migrate to the center \citep{Noguchi1999,Immeli2004}.
More generally, the gas in the disk can fall to the center in a dissipative way, an event called wet compaction \citep{DekelBurkert2014,Zolotov2015,Tacchella2016}.
All these paths predict a phase in which star formation is concentrated in the central kiloparsecs and is heavily obscured by dust.

ALMA observations of massive star-forming galaxies at $z\sim2$ support this picture.
The 870~$\mu$m dust continuum, which traces the obscured star formation, is compact, with typical effective radii of 1--2~kpc \citep{Simpson2015,Ikarashi2015,Hodge2016, Gullberg2019,Ikeda2022}.
High-resolution observations of submillimeter galaxies first revealed that the dust and gas emission comes from regions that are faint in the rest-frame UV and optical \citep{Hodge2012,Chen2015}.
The dust emission appeared much more compact than the stellar light measured in the rest-frame UV and optical with HST \citep{Barro2016,Lang2019,Tadaki2020}, and this contrast was taken as evidence for a compact starburst that builds a bulge inside a larger stellar disk.
This interpretation has a weakness.
The rest-frame UV and optical light is strongly affected by dust attenuation, in particular in the obscured centers \citep{Popping2022}, so the true stellar mass distribution of these galaxies could not be measured reliably.

JWST has changed this situation.
NIRCam imaging showed that massive galaxies at $1\leq z\leq2.5$ are smaller in the rest-frame near-infrared than in the rest-frame optical \citep{Suess2022}, and stellar bulges were found in submillimeter galaxies at 4.4~$\mu$m \citep{Chen2022}.
Multi-band NIRCam analyses added a general lesson about the colors of massive galaxies.
Massive star-forming galaxies at these redshifts are more attenuated than earlier estimates suggested \citep{vanderWel2025}, and they show strongly negative color gradients, with centers much redder than the outskirts, that are driven mostly by radial gradients of the dust attenuation \citep{Price2025,Martorano2026}.
The stellar mass distribution is therefore more concentrated than the light distribution even in the near-infrared bands.
MIRI imaging extends this progress to longer wavelengths \citep{Boogaard2024,Lyu2025,Li2026}.
At $z\sim2$, the F770W band probes the rest-frame 2.4~$\mu$m continuum, which corresponds to the rest-frame $K$ band.
The rest-frame $K$-band light is dominated by the evolved stellar population and is much less affected by dust attenuation than shorter wavelengths, and it has long been the standard tracer of the stellar mass distribution in the local universe.
MIRI F770W imaging therefore gives a rest-frame $K$-band view of the stellar mass distribution in these dusty galaxies, at a resolution of $0\farcs28$ (FWHM).

In this paper, we combine JWST MIRI F770W imaging, NIRCam imaging in eight bands, and ALMA 870~$\mu$m data for 16 massive dusty star-forming galaxies at $z\sim2$ drawn from \citet{Tadaki2020}.
We first measure the F770W sizes of the individual galaxies and compare them with the sizes of the dust emission.
We then stack the 16 galaxies in the ten bands at a matched resolution, decompose the stacked F770W profile into a bulge and a disk, and fit the SEDs in radial bins with CIGALE to derive the radial profiles of the stellar mass, the SFR, and the dust attenuation.
Throughout this paper, we adopt a flat $\Lambda$CDM cosmology with $H_0=70~\mathrm{km~s^{-1}~Mpc^{-1}}$ and $\Omega_\mathrm{m}=0.3$, and a \citet{Chabrier2003} initial mass function.

\section{Sample and Data} \label{sec:data}

\subsection{Sample} \label{sec:sample}

Our targets are drawn from the ALMA 870~$\mu$m survey of massive star-forming galaxies at $z\sim2$ presented in \citet{Tadaki2020}.
\citet{Tadaki2020} observed 85 massive ($\log(M_\star/\Msun)>11$) galaxies at $1.9<z<2.6$ in the CANDELS UDS and GOODS-S fields, selected from the 3D-HST catalogs \citep{Grogin2011,Skelton2014,Momcheva2016}.
With the rest-frame UVJ colors, 78 of them are classified as star-forming galaxies, which form a mass-selected sample of massive star-forming galaxies.
The 870~$\mu$m continuum is detected for 73 of the 78 galaxies (94\%) at $\mathrm{S/N}>5$, so essentially all massive star-forming galaxies at this epoch are dusty.
Our analysis requires a reliable size measurement of the dust emission.
We therefore start from the 62 galaxies (79\%) detected at $\mathrm{S/N}>10$, for which \citet{Tadaki2020} measured the sizes of the dust emission through visibility-based modeling.
Among these 62 galaxies, 16 are covered by the MIRI F770W imaging of the PRIMER UDS survey (Section~\ref{sec:jwst}).
This last step depends only on the MIRI coverage and not on the properties of the galaxies, so it does not introduce a bias.
They constitute the sample of this paper and are listed in Table~\ref{tab:sample}.
The 16 galaxies span a redshift range of $z=1.95$--2.53 with a median of $z_\mathrm{med}=2.18$, stellar masses of $\log(M_\star/\Msun)=11.0$--11.5 (median 11.2), and star formation rates of $\mathrm{SFR}\simeq180$--$1400~\Msun~\mathrm{yr^{-1}}$ (median $\simeq300~\Msun~\mathrm{yr^{-1}}$), based on the infrared luminosities \citep{Tadaki2020}.
Our targets are thus not a specially dusty subset of the population but are representative of the massive end of the star-forming main sequence at this epoch.

\begin{deluxetable}{lcccccc}
\tablecaption{The sample of 16 massive star-forming galaxies
\label{tab:sample}}
\tablehead{
\colhead{ID} & \colhead{$z$} & \colhead{$S_{870}$} &
\colhead{$R_\mathrm{e,c}$(870~$\mu$m)} & \colhead{$m_\mathrm{F770W}$} &
\colhead{$R_\mathrm{e,c}$(F770W)} & \colhead{$R_\mathrm{e,c}$(F444W)} \\
\colhead{} & \colhead{} & \colhead{(mJy)} &
\colhead{(kpc)} & \colhead{(AB mag)} & \colhead{(kpc)} & \colhead{(kpc)}
}
\startdata
U4-190 & 2.06 & 5.64 $\pm$ 0.14 & 0.67 $\pm$ 0.03 & 20.03 & 0.67 $\pm$ 0.01 & 1.04 \\
U4-394 & 2.28 & 1.53 $\pm$ 0.18 & 1.06 $\pm$ 0.26 & 20.93 & 1.24 $\pm$ 0.01 & 1.64 \\
U4-1833 & 2.51 & 1.94 $\pm$ 0.16 & 2.33 $\pm$ 0.49 & 20.85 & 2.07 $\pm$ 0.02 & 2.18 \\
U4-4059 & 2.323 & 1.49 $\pm$ 0.20 & 0.49 $\pm$ 0.13 & 20.77 & 0.98 $\pm$ 0.01 & 1.39 \\
U4-7472 & 2.093 & 3.16 $\pm$ 0.19 & 1.78 $\pm$ 0.23 & 20.90 & 1.57 $\pm$ 0.02 & 1.68 \\
U4-7516 & 2.08 & 2.37 $\pm$ 0.24 & 2.87 $\pm$ 0.41 & 20.42 & 2.63 $\pm$ 0.02 & 2.70 \\
U4-34454 & 2.027 & 1.63 $\pm$ 0.13 & 2.47 $\pm$ 0.48 & 20.48 & 1.78 $\pm$ 0.03 & 2.59 \\
U4-34617 & 2.53 & 2.21 $\pm$ 0.18 & 0.31 $\pm$ 0.20 & 21.28 & 1.76 $\pm$ 0.03 & 2.47 \\
U4-34817 & 2.19 & 2.02 $\pm$ 0.19 & 3.22 $\pm$ 0.64 & 20.62 & 3.10 $\pm$ 0.03 & 3.41 \\
U4-36247 & 2.179 & 1.80 $\pm$ 0.19 & 0.52 $\pm$ 0.12 & 20.96 & 1.76 $\pm$ 0.03 & 2.24 \\
U4-36437 & 2.083 & 2.01 $\pm$ 0.18 & 0.61 $\pm$ 0.12 & 20.92 & 0.87 $\pm$ 0.01 & 1.29 \\
U4-36568 & 2.177 & 1.26 $\pm$ 0.21 & 2.80 $\pm$ 0.70 & 21.05 & 2.42 $\pm$ 0.03 & 2.56 \\
U4-36685 & 1.95 & 1.60 $\pm$ 0.17 & 1.32 $\pm$ 0.38 & 19.84 & 0.54 $\pm$ 0.01 & 1.92 \\
U4-40115 & 2.02 & 2.18 $\pm$ 0.14 & 0.73 $\pm$ 0.14 & 20.98 & 1.04 $\pm$ 0.01 & 1.31 \\
U4-42529 & 2.39 & 2.56 $\pm$ 0.18 & 2.06 $\pm$ 0.34 & 20.36 & 1.94 $\pm$ 0.01 & 2.01 \\
U4-42571 & 2.41 & 2.10 $\pm$ 0.17 & 1.54 $\pm$ 0.25 & 20.05 & 0.61 $\pm$ 0.01 & 1.32 \\
\enddata
\tablecomments{
IDs and redshifts are taken from \citet{Tadaki2020} and are based on the
3D-HST catalogs \citep{Skelton2014}.
$S_{870}$ is the 870~$\mu$m flux density from Table~1 of
\citet{Tadaki2020}.
$R_\mathrm{e,c}$(870~$\mu$m) is the circularized effective radius of the
870~$\mu$m emission, $R_\mathrm{e,c}=R_\mathrm{e}\sqrt{q}$, computed
from the effective radius $R_\mathrm{e}$ of the exponential-disk fit and
the axis ratio $q$ of the elliptical Gaussian fit in Table~3 of
\citet{Tadaki2020}.
$m_\mathrm{F770W}$ is the total AB magnitude of the F770W emission from
the $n=1$ fit in this work (Section~\ref{sec:sizes}). Its statistical
uncertainty is 0.01--0.02~mag.
$R_\mathrm{e,c}$(F770W) is the circularized effective radius measured in
this work with an exponential-disk fit to the MIRI F770W images
(Section~\ref{sec:sizes}).
$R_\mathrm{e,c}$(F444W) is the circularized effective radius of the
F444W emission, from the $n=1$ fits of
\citet{Tadaki2023}.
}
\end{deluxetable}

\subsection{JWST Imaging} \label{sec:jwst}

The CANDELS UDS field was observed with JWST/NIRCam and MIRI as part of the PRIMER survey (GO-1837, PI J.~Dunlop, \citealt{Donnan2024}).
We use the publicly available mosaics from the DAWN JWST Archive (DJA)\footnote{\url{https://dawn-cph.github.io/dja/}}, version 7.2, which were processed with the \texttt{grizli} pipeline (\citealt{Brammer2023grizli}; see \citealt{Valentino2023} for a description of the image processing).
All mosaics are drizzled onto a common $0\farcs04$ pixel grid.

The MIRI F770W imaging covers only part of the field, in two mosaics placed in the northern and southern parts of it, and all 16 targets fall on one of the two mosaics.
The on-source exposure times at the positions of our targets are 1.8--5.3~ks in F770W.
At the median redshift of the sample, the F770W band probes the rest-frame $\sim$2.4~$\mu$m ($K$-band) continuum, which traces the stellar mass distribution.
In addition to F770W, we use imaging in eight NIRCam bands (F090W, F115W, F150W, F200W, F277W, F356W, F410M, and F444W), which sample the rest-frame UV to near-infrared ($\sim$0.3--1.4~$\mu$m) continuum, for the stacking and spatially resolved SED analyses (Sections~\ref{sec:stacking} and \ref{sec:sed}).

\subsection{ALMA Data} \label{sec:alma}

All 16 galaxies were observed in the ALMA Band~7 continuum ($\lambda\simeq870~\mu$m) with both a compact and an extended array configuration.
The compact-configuration data are taken from project 2017.1.01027.S \citep{Tadaki2020} for all targets.
The extended-configuration data are also from 2017.1.01027.S except for four galaxies (U4-34617, U4-34817, U4-36247, and U4-36568), for which we use the extended-configuration data from project 2012.1.00245.S \citep{Tadaki2017}.
The correlator was configured with four spectral windows in two sidebands covering 335.5--351.5~GHz. The tuning of the 2012.1.00245.S observations is offset by $\simeq$1.5~GHz from that of 2017.1.01027.S.
The combined visibilities continuously sample projected baselines from $\sim$10 to 1800~k$\lambda$ without a gap between the two configurations, which is important for the quality of the synthesized beam of the stacked image (Section~\ref{sec:stacking}).
We start our analysis from the calibrated measurement sets. The observations and calibration are described in detail in \citet{Tadaki2017} and \citet{Tadaki2020}.

\section{F770W Sizes of Individual Galaxies} \label{sec:sizes}

In this section we measure the effective radii of the 16 galaxies in the F770W band and compare them with the 870~$\mu$m dust sizes of \citet{Tadaki2020} and with the F444W sizes of \citet{Tadaki2023}.
The FWHM of the F770W point-spread function (PSF), $0\farcs28$, corresponds to 2.3~kpc at $z=2.18$ and is comparable to the expected galaxy sizes. The size measurements therefore depend on an accurate PSF model, which we construct first.

\subsection{F770W PSF} \label{sec:psf}

The MIRI PSF shows six diffraction spikes and the cruciform artifact characteristic of the short-wavelength MIRI bands \citep{Gaspar2021,Wright2023}.
Both features are fixed in the detector frame and rotate on the sky with the telescope roll angle.
From the exposure information of the mosaics and the observation footprints registered in the Mikulski Archive for Space Telescopes (MAST), we find that the JWST observations contributing to each mosaic were taken at sky roll angles that agree within $\sim$$6\arcdeg$, and that each target is covered at essentially a single roll angle among three discrete values.
We therefore build a single high signal-to-noise PSF, rotate it to the roll angle appropriate for each target, and combine the rotated PSFs with exposure-time weighting for the few targets that lie in overlap regions of observations with different roll angles.

\begin{figure}
\centering
\figfile[scale=1.0]{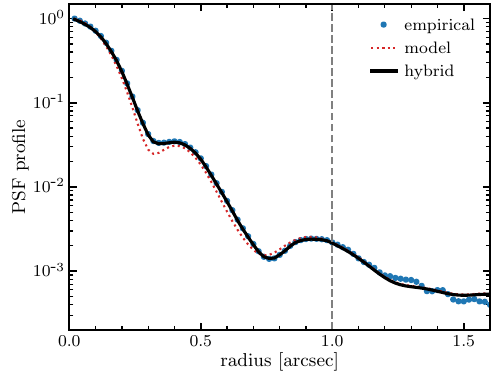}
\caption{
Azimuthally averaged radial profiles of the F770W PSF models, normalized at the peak.
The empirical PSF (circles) is the median stack of 21 Gaia-selected stars, and the dotted curve shows the \texttt{STPSF} model.
The adopted hybrid PSF (solid curve) follows the empirical stack at $r\leq 1\farcs0$ (dashed vertical line) and the rescaled model wing at larger radii.
}
\label{fig:psf}
\end{figure}

We construct a hybrid PSF that combines an empirical stellar stack in the core with the \texttt{STPSF} model \citep{Perrin2014} in the wings.
For the empirical part, we select PSF stars from Gaia~DR3 \citep{GaiaDR3}, whose parallaxes and proper motions provide a secure star--galaxy separation that is otherwise difficult at the F770W resolution.
After requiring full coverage, accurate centering, isolation, and an unsaturated core, and after a visual screening that removes a close binary, 21 stars are retained.
Each star is background-subtracted, masked for neighboring sources, recentered with subpixel accuracy, and normalized, and the stars are median-stacked on a $0\farcs02$ grid, twice finer than the mosaic pixels.
The stacked PSF has a measured $\mathrm{FWHM}$ of $0\farcs281$.
For the model part, we generate an \texttt{STPSF} PSF including the detector effects, convolve it with the pixel and drizzle kernel responses, and resample it onto the same grid.
The absolute orientation of the model is calibrated against the stellar stack through a rotational cross-correlation, which fixes the zero point of the roll angle. After this calibration, the residual misalignment of the diffraction spikes is consistent with zero.
The model agrees well with the stellar stack at large radii, but its core is 6--7\% sharper ($\mathrm{FWHM}=0\farcs263$) than observed (Figure~\ref{fig:psf}).
We therefore adopt the empirical stack at $r\leq1\farcs0$ and the model wing at larger radii, rescaled to make the profile continuous at the splice radius, and normalize the combined PSF to unit total flux.

\subsection{S\'ersic Fits and Uncertainties} \label{sec:galfit}

\begin{figure*}
\centering
\figfile[scale=1.0]{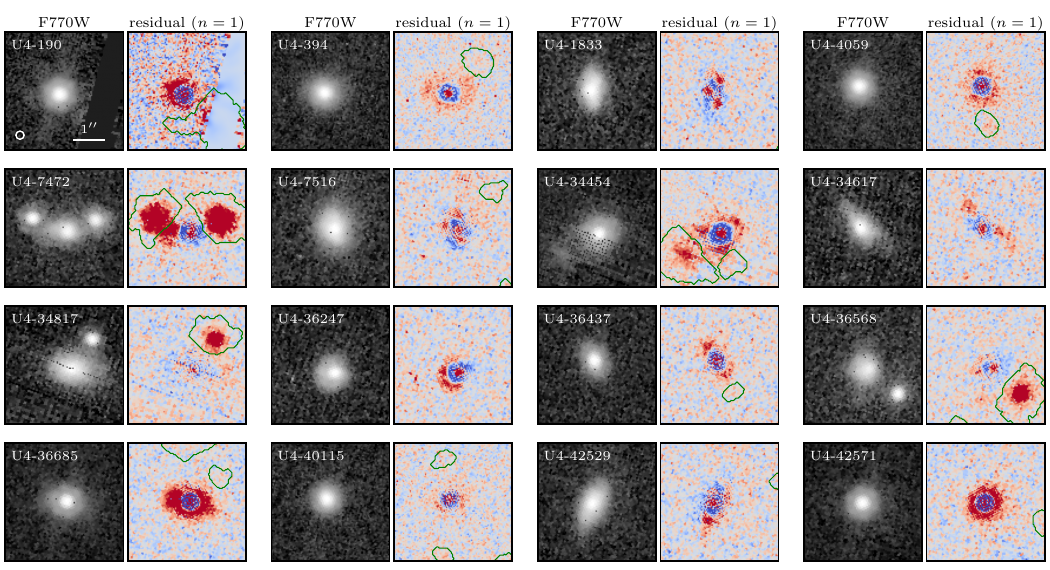}
\caption{
F770W images and GALFIT residuals of the 16 galaxies.
For each galaxy, the left panel shows the F770W cutout ($4''$ on a side, arcsinh stretch) and the right panel shows the residual of the $n=1$ fit on a linear scale from $-5\sigma$ to $+5\sigma$.
The green contours show the neighbor masks built from the F444W images.
The white circle and the white bar in the first panel show the PSF FWHM ($0\farcs281$) and a length of $1''$.
}
\label{fig:cutouts}
\end{figure*}

We fit two-dimensional S\'ersic models to $6''$ cutouts with GALFIT \citep{Peng2002,Peng2010}, using the hybrid PSF assigned to each target at twice-finer sampling than the science images.
The noise images are derived from the weight maps of the mosaics including the Poisson contribution of the sources.
Neighboring objects are masked using segmentation maps constructed from the deeper and sharper F444W images. The same masks are used for all the JWST bands in the stacking analysis (Section~\ref{sec:stacking}).
Figure~\ref{fig:cutouts} shows the F770W cutouts of the 16 galaxies together with the residuals of the fits described below and the masks.

Each galaxy is fit with a S\'ersic $+$ constant-sky model.
The S\'ersic index is fixed to $n=1$ for a direct comparison with the 870~$\mu$m sizes, which were measured with an exponential-disk model \citep{Tadaki2020}.
The free parameters are the central position, the total magnitude, the effective radius $R_\mathrm{e}$, the axis ratio $q$, the position angle, and the sky level.
Fits with the S\'ersic index also free, within an allowed range of $n=0.2$--8.0, yield a median $n=2.4\pm0.5$, indicating centrally concentrated profiles (Section~\ref{sec:bulgedisk}), but change the sizes only mildly in a systematic sense. The median ratio of the free-$n$ to $n=1$ circularized radii is $1.01\pm0.03$.
Here and in the following, the error of a median value is the bootstrap uncertainty of the median, obtained by resampling the 16 galaxies with replacement.
We quote circularized effective radii, $R_\mathrm{e,c}=R_\mathrm{e}\sqrt{q}$, from the $n=1$ fits throughout.

Because the drizzled mosaics have strongly correlated pixel noise, the formal uncertainties reported by GALFIT are not reliable.
We instead estimate the uncertainties with an empty-sky perturbation method. For each galaxy, we add to the real cutout each of 50 source-free sky cutouts drawn from regions of the same mosaic with matching exposure time, refit the perturbed images, and adopt half of the 16th--84th percentile range of the output parameters as the 1$\sigma$ uncertainty.
The resulting uncertainties of $R_\mathrm{e,c}$ ($n=1$) are typically 1\% (Table~\ref{tab:sample}).
They do not include the systematic uncertainty associated with the PSF model, which we assess in Appendix~\ref{app:calib} with injection and recovery simulations using two independent subsets of the PSF stars.
The simulations show that the PSF-related bias is 1 to 2\% for the sizes measured in our sample.

\subsection{Comparison of the F770W, 870~$\mu$m, and F444W Sizes}
\label{sec:sizecomp}

\begin{figure*}
\centering
\figfile[scale=1.0]{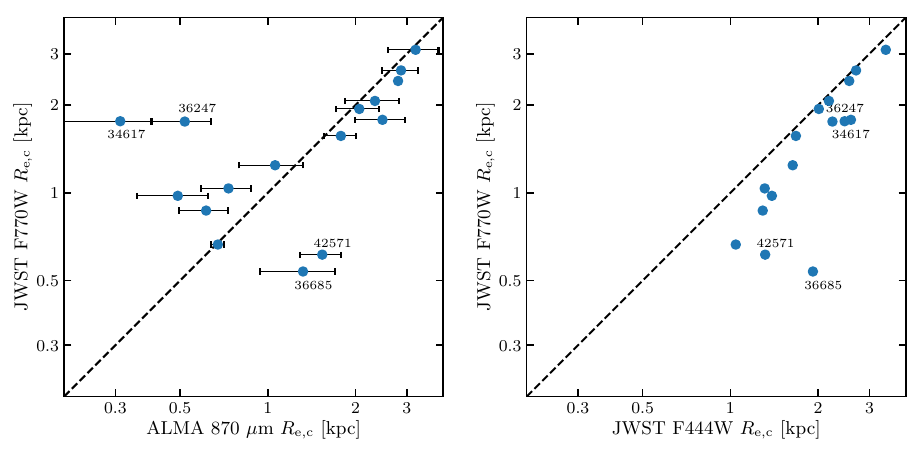}
\caption{
Circularized effective radii of the F770W emission (this work, $n=1$) compared with those of the 870~$\mu$m dust continuum \citep[][left panel]{Tadaki2020} and those of the F444W emission \citep[][right panel]{Tadaki2023} for the 16 galaxies.
The dashed lines show the one-to-one relations.
The F770W error bars are the statistical uncertainties from the empty-sky perturbation analysis (Section~\ref{sec:galfit}) and are mostly smaller than the symbols.
The four outliers discussed in the text are labeled.
}
\label{fig:size_alma}
\end{figure*}

The left panel of Figure~\ref{fig:size_alma} compares the circularized effective radii measured in F770W with those of the 870~$\mu$m dust continuum.
The two sizes track each other well over the full range of $R_\mathrm{e,c}\simeq0.5$--3~kpc, with a median ratio of $R_\mathrm{e,c}(\mathrm{F770W})/R_\mathrm{e,c}(870~\mu\mathrm{m})=0.95\pm0.13$.
The scatter around the one-to-one relation is dominated by the 870~$\mu$m measurement uncertainties.
The centroids of the two emissions also coincide. After removing a small systematic astrometric offset between the ALMA and JWST frames ($\simeq$$0\farcs02$), the median separation between the 870~$\mu$m and F770W centers is $0\farcs038$ (0.31~kpc), with a maximum of $0\farcs11$.
Thus, the dust continuum and the rest-frame $K$-band stellar light are cospatial and have a comparable spatial extent.

Four galaxies deviate clearly from the one-to-one relation.
In U4-34617 and U4-36247, the 870~$\mu$m emission is 3 to 6 times more compact than the F770W emission.
For U4-34617, the compact-disk model of \citet{Tadaki2020} recovers only about half of the flux measured in the low-resolution ALMA image, so this galaxy also hosts an extended dust component and the fitted size describes only the central part.
For U4-36247, the compact disk recovers almost all of the 870~$\mu$m flux.
In both galaxies, an extended stellar disk coexists with a more concentrated dusty star-forming region.
Such a two-component structure is expected in the wet compaction scenario, in which gas falls to the center in a dissipative way and feeds a compact starburst \citep[e.g.,][]{DekelBurkert2014, Zolotov2015}.
In U4-42571 and U4-36685, on the other hand, the F770W emission is more compact than the 870~$\mu$m emission.
These are the only two galaxies whose free-$n$ fits reach the upper bound of $n=8$, which points to a strong point-like component at the center.
The residual maps of these two galaxies in Figure~\ref{fig:cutouts} show a clear ring-like pattern, which also means that a single exponential profile cannot describe their centers.
Warm dust heated by an AGN can produce such a red and point-like nucleus at rest-frame 2.4~$\mu$m.
The $n=1$ sizes of these two galaxies should therefore be taken as lower limits on the extent of their stellar disks.

The F770W sizes are also systematically smaller than the F444W sizes measured with $n=1$ fits by \citet{Tadaki2023}, which are listed in Table~\ref{tab:sample} (right panel of Figure~\ref{fig:size_alma}).
All 16 galaxies have $R_\mathrm{e,c}(\mathrm{F770W})<R_\mathrm{e,c}(\mathrm{F444W})$, with a median ratio of $0.77\pm0.07$ (range 0.28--0.97).
The stellar light at rest-frame 2.4~$\mu$m is thus systematically more compact than at rest-frame 1.4~$\mu$m.
This implies a negative radial color gradient, with redder centers, as expected if the dust attenuation or the stellar mass-to-light ratio increases toward the center.
We quantify these gradients with the stacked multi-band profiles and the spatially resolved SED analysis in Sections~\ref{sec:stacking} and \ref{sec:sed}.

\section{Stacking Analysis} \label{sec:stacking}

The individual galaxies are detected at high significance only in their central regions.
To trace the radial distributions of the stellar light and the dust emission out to larger radii, and to construct spatially resolved SEDs (Section~\ref{sec:sed}), we stack the 16 galaxies in ten bands, the nine JWST bands and the ALMA 870~$\mu$m continuum.
All stacks share a common design.
Each galaxy enters every band normalized by its own total F770W flux from the $n=1$ fit ($m_\mathrm{F770W}$ in Table~\ref{tab:sample}).
The stacks are centered on the F770W centroids, each galaxy carries equal weight, and the mean of the 16 galaxies is taken.
Finally, all ten stacks are matched to a common angular resolution, the PSF of the F770W stack ($\mathrm{FWHM}=0\farcs28$).

\subsection{JWST Stacks} \label{sec:jwststack}

For each JWST band, we extract cutouts on the same pixel grid as in Section~\ref{sec:galfit}, subtract a local sky, mask neighboring objects with the F444W-based segmentation masks, align the images on the F770W centroids on a twice-oversampled ($0\farcs02$) grid, and normalize them by the F770W fluxes.
A hybrid PSF is constructed for each NIRCam band with the same procedure as for F770W (Section~\ref{sec:psf}).
The number of usable PSF stars is 13 to 40 per band.
Each band is then convolved to the F770W stack PSF with a matching kernel constructed with \texttt{photutils} \citep{Bradley2026}.
The accuracy of the kernels, measured as the residual of the convolved PSF relative to the target PSF, is 0.6--2.5\% of the total flux.
The masked images are mean-combined, and radial profiles are extracted in circular annuli of $0\farcs04$ width.
We use the profiles out to $r=1\farcs5$ (38 annuli) throughout this paper.

We evaluate two error components as full radius--radius covariance matrices.
The noise covariance is obtained by repeating the identical stacking procedure on 500 sets of source-free sky positions matched in field and exposure time.
The sample covariance, which accounts for the galaxy-to-galaxy variance, is obtained with a paired bootstrap. The 16 galaxies are resampled with replacement 50 times, and an identical set of bootstrap indices is used in all ten bands, so that the correlations between bands are preserved and propagate consistently into the colors and the resolved SEDs.
The total uncertainty is the sum of the two covariances.
Within the adopted limit of $r=1\farcs5$, the sample term dominates.
At larger radii the noise term takes over.

\subsection{ALMA 870~$\mu$m Stack} \label{sec:almastack}

Before the stacking, we removed the 870~$\mu$m emission of neighboring sources.
In five galaxies (U4-394, U4-4059, U4-7472, U4-34817, and U4-36568), the ALMA data detect 870~$\mu$m emission from sources other than the target.
Following \citet{Tadaki2020}, we fit each neighboring source, together with the target, in the visibility plane with \textsc{uvmultifit} \citep{MartiVidal2014}, assuming a circular Gaussian model for the neighboring source, and obtained its position, flux density, and effective radius.
We then subtracted the best-fit model of the neighboring source from the visibilities.

The ALMA data are stacked in the $uv$ plane with CASA \citep{CASA2022}.
Stacking in the $uv$ plane rather than in the image plane keeps the visibilities after the stacking and gives more robust size estimates for resolved sources \citep{Lindroos2015}.
Each calibrated measurement set is phase-shifted to the F770W centroid of the target after removing the small median offset between the ALMA and JWST astrometric frames ($\simeq$$0\farcs02$, Section~\ref{sec:sizecomp}).
We verified that stacking on the 870~$\mu$m centers instead gives almost the same profiles.

The stack is then assembled with three operations on the per-galaxy measurement sets.
First, the complex visibilities (the \texttt{DATA} column) of each galaxy are divided by its total F770W flux, the same per-galaxy normalization as in the JWST stacks.
Second, the visibility weights (the \texttt{WEIGHT} column, computed with the CASA task \texttt{statwt}) are adjusted in two steps.
The weights of the compact-configuration data are first multiplied by a factor of 1/3.
Without this down-weighting, the dense short-baseline coverage of the compact configuration would produce a dirty beam that looks like a superposition of two Gaussians, with a narrow core on top of a broad base.
The down-weighting keeps the dirty beam close to a single Gaussian.
The weights of each galaxy are then rescaled by a single factor such that their sum equals unity.
This rescaling preserves the relative weights within each data set, while forcing every galaxy to contribute exactly the same total weight to the stack.
The 870~$\mu$m stack is therefore an equal-weight mean of the 16 normalized galaxies, one vote per galaxy, which is the direct $uv$-plane analog of the equal-weight mean of the flux-normalized images in the JWST stacks.
Third, the phase-center directions recorded in the \texttt{FIELD} tables of the 16 measurement sets are overwritten with a single common direction, and the data sets are concatenated.

\begin{figure*}[!t]
\centering
\figfile[scale=1.0]{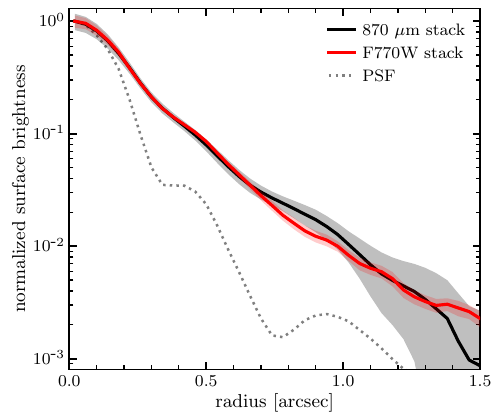}\hfill
\figfile[scale=1.0]{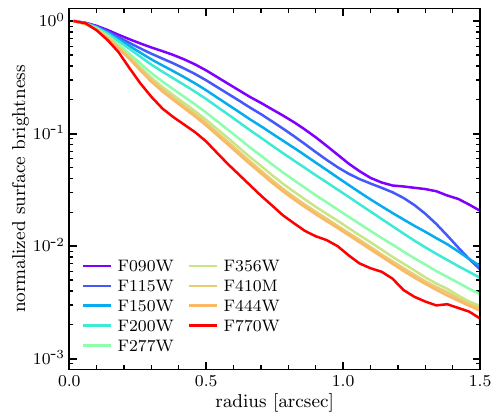}
\caption{
Radial surface-brightness profiles of the stacked emission, normalized at the peak.
Left panel. The F770W stack (red) and the 870~$\mu$m stack (black).
The shaded bands show the total 1$\sigma$ uncertainties (noise $+$ sample), and the dotted curve shows the PSF of the F770W stack, to which all the bands are matched.
The two profiles agree closely out to $r\simeq1''$ ($\simeq$8~kpc).
Right panel. The profiles of the nine JWST bands.
The profiles become more compact toward longer wavelengths, which shows the radial color gradient of the stacked light.
}
\label{fig:profiles}
\end{figure*}

The stacked visibilities are imaged with multi-scale \textsc{clean} \citep{Cornwell2008}, and the cleaning is restricted to a circular mask of $1\farcs5$ radius centered on the stacked source.
The synthesized beam must be smaller than the F770W PSF for the subsequent resolution matching, while retaining as much short-baseline sensitivity to extended emission as possible.
We achieve this with the down-weighting of the compact configuration introduced above, combined with Briggs weighting with $\mathrm{robust}=0.7$, which yields a $0\farcs244\times0\farcs216$ beam.
Because of the gap-free baseline coverage (Section~\ref{sec:alma}), the dirty beam is nearly Gaussian, with sidelobes below 10\% of the peak.
The residual image is scaled by the ratio of the clean-beam to dirty-beam volumes ($\epsilon=0.47$) before being added back, following \citet{Jorsater1995} \citep[see][]{Czekala2021}, and the corrected image is convolved to the F770W stack PSF with a matching kernel constructed from the elliptical Gaussian clean beam (accuracy $\sim$1\%).

The uncertainties mirror the JWST ones.
The sample covariance is computed from the same 50 bootstrap realizations with the same shared indices, where each realization is re-imaged with the full pipeline, including the Briggs re-weighting, the multi-scale clean, the residual scaling, and the kernel convolution.
The noise covariance is computed with a sign-flip jackknife.
In each of 50 trials, a random half of the visibilities is sign-flipped and the data are re-imaged, which removes the sky signal while keeping the noise statistics.

\subsection{Radial Profiles of the Stacked Emission}
\label{sec:profiles}

The left panel of Figure~\ref{fig:profiles} shows the radial profiles of the F770W and 870~$\mu$m stacks, normalized at the peak, together with the F770W stack PSF.
Both stacks are well resolved, and their profiles are remarkably similar over the whole radial range where both are detected at high significance ($r\lesssim1\farcs0$, corresponding to $\simeq$8~kpc).
This extends the result of Section~\ref{sec:sizecomp} from a comparison of effective radii to the full profile shape.
The dust continuum, which traces the obscured star formation, follows the radial distribution of the rest-frame $K$-band stellar light.

The right panel of Figure~\ref{fig:profiles} shows the profiles of the nine JWST bands.
The profiles become more compact toward longer wavelengths in a smooth and ordered way, from F090W to F770W.
This wavelength dependence is the radial color gradient behind the size trend of Section~\ref{sec:sizecomp}, and it is now traced continuously over the whole radial range.
The centers are redder in every color, as expected when the dust attenuation increases toward the center.
Even F444W is clearly more extended than F770W, so the light at rest-frame 1.4~$\mu$m is still affected by the dust attenuation in the central region.

Two features of the bluest bands deserve a note.
In the individual galaxies, the F090W images, which probe the rest-frame UV, show clumpy structures, and bright clumps often appear at $r\sim0\farcs5$ rather than in the central region.
A similar behavior is clearly seen in GN20, a starburst galaxy at $z=4$ \citep{Boogaard2026}.
Unobscured star-forming clumps therefore contribute to the F090W profile around these radii.
At larger radii ($r>1''$), the decline of the F090W and F115W profiles becomes much shallower, but this reflects the low signal-to-noise ratio of the faint outer emission in the bluest bands rather than a real structure.
The colors among the ten stacked bands, and their radial dependence, are analyzed through the spatially resolved SEDs in Section~\ref{sec:sed}.

\subsection{Bulge--Disk Decomposition of the F770W Stack}
\label{sec:bulgedisk}

We first fit the radial profile of the F770W stack with a single S\'ersic model, in which the flux, the effective radius, and the S\'ersic index are all free, to check that the stack represents the sample.
Two-dimensional circular models are rendered on the $0\farcs02$ grid, convolved with the F770W stack PSF, binned to the mosaic pixel scale, and their profiles are extracted with the same annuli as the data.
The free parameters are optimized by minimizing a $\chi^2$ that uses the full total covariance matrix of the profile out to $r=1\farcs5$, inside which sample variance rather than sky noise dominates the error budget (Section~\ref{sec:jwststack}), so that the result is not driven by the noisier outer bins.
All the confidence intervals of the fitted parameters in this section are the 16--84\% ranges of the 50 bootstrap realizations of the profile, which are refitted with a fixed weight matrix.
The fit gives $n=2.6$ (2.0--2.9) and $R_\mathrm{e,c}=1.5$~kpc (1.4--1.8~kpc), where the model is circular so that its effective radius is directly a circularized radius.
Both values agree with the medians of the individual free-$n$ fits, $n=2.4\pm0.5$ (Section~\ref{sec:galfit}) and $R_\mathrm{e,c}=1.9\pm0.4$~kpc.
The stacked profile therefore represents the typical galaxy of the sample, and it is not driven by a subset of the galaxies.

\begin{figure}
\centering
\figfile[scale=1.0]{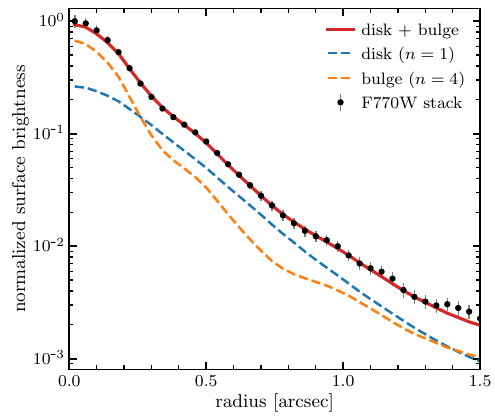}
\caption{
Bulge--disk decomposition of the radial profile of the F770W stack.
The left panel shows the stacked profile (points with total 1$\sigma$ uncertainties) and the best-fit PSF-convolved model (solid curve), which consists of an exponential disk ($n=1$, blue dashed) and a de Vaucouleurs bulge ($n=4$, orange dashed) with $B/T=0.50$.
The right panel shows the residuals normalized by the total uncertainty.
}
\label{fig:bulgedisk}
\end{figure}

The single S\'ersic model, however, leaves systematic residuals (reduced $\chi^2=3.6$), and its index lies between an exponential disk and a classical bulge, which suggests a composite structure.
We therefore decompose the radial profile of the F770W stack into an exponential disk ($n=1$) and a de Vaucouleurs bulge ($n=4$) in the same way.
The four free parameters are the flux and effective radius of each component.
The best-fit model gives $R_\mathrm{e,disk}=2.43$~kpc (2.40--2.48~kpc), $R_\mathrm{e,bulge}=0.82$~kpc (0.61--0.93~kpc), and a bulge-to-total ratio of $B/T=0.50$ (0.43--0.54) at rest-frame 2.4~$\mu$m (Figure~\ref{fig:bulgedisk}).
The fit leaves coherent residuals at the level of reduced $\chi^2=2.0$ (34 degrees of freedom), indicating that the average profile is not perfectly described by the two-component model.
The two-component model is nevertheless a much better description than the single S\'ersic model.

We tested whether the compact component is resolved by fitting two alternative models with the same data and weight matrix.
Replacing the bulge with a point source (a point source plus an $n=1$ disk, three free parameters) degrades the fit substantially, from $\chi^2=69$ to 98, and is rejected by the information criteria ($\Delta\mathrm{BIC}=25$).
The central light excess over the disk is broader than the PSF.
Adding a point source to the disk$+$bulge model (five free parameters) improves $\chi^2$ only marginally (from 69 to 66), which does not justify the extra parameter ($\Delta\mathrm{BIC}=+0.5$ with respect to the disk$+$bulge model), and the point-source flux converges to only $\simeq$6\% of the total.
The disk plus resolved bulge decomposition is therefore the preferred description of the stacked profile, and any unresolved nuclear component (for example an AGN) contributes at most a few percent of the rest-frame 2.4~$\mu$m light.

Two caveats apply.
First, the bulge effective radius is only $\sim$35\% of the PSF FWHM, so its exact value is sensitive to the accuracy of the PSF core and is partially degenerate with $B/T$.
Second, the stack is an average of 16 galaxies with different sizes and shapes, and such a mixture can itself produce a profile that looks composite.
The decomposition should therefore be taken as a description of the light distribution of a representative galaxy, not as a structural measurement of any single system.
With these caveats, the decomposition agrees with the picture from the individual measurements.
About 50\% of the rest-frame 2.4~$\mu$m light is in a compact ($\simeq$0.8~kpc) component, which is embedded in an extended disk.

An alternative interpretation of the compact component is a stellar bar.
A similar two-component structure was found in the stacked 870~$\mu$m profile of about 700 submillimeter galaxies \citep{Gullberg2019}, and the compact dust component has been linked to bar-like structures rather than to a bulge \citep{Hodge2019}.
Because the stacking erases the azimuthal information, our circular decomposition cannot distinguish a bulge from a bar.
In either case, the compact component represents a dense stellar structure in the central kiloparsec, a bulge or its precursor.

\section{Spatially Resolved SED Fitting} \label{sec:sed}

\subsection{SEDs in Radial Bins} \label{sec:sedinput}

We now combine the ten stacked profiles into SEDs in radial bins and fit them with CIGALE.
The first bin is the central resolution element, for which we use the central pixel alone.
It is a PSF-weighted average of the intrinsic profile, with half of the weight coming from $r\lesssim0\farcs14$.
The fine annuli at $r=0\farcs04$--$0\farcs16$ lie inside this central resolution element and carry no independent information, so they are not used.
The following bins are annuli with a width of $0\farcs28$, which equals the FWHM of the F770W stack PSF, starting at $r=0\farcs16$.
The fine annuli inside each bin are averaged with area weights.
The averaging is done with a fixed matrix, and the same matrix is applied to the central profiles, to every bootstrap realization, and to the covariance matrices.
In this way, the correlated errors of the annuli inside each bin are propagated correctly.
We fit the central resolution element and the four inner annuli, which cover $r=0$ to $1\farcs28$ (10.6~kpc).

The normalized profiles are placed on an absolute flux scale with a single number.
All bands and all bins are multiplied by the median F770W total flux of the 16 galaxies (17.2~$\mu$Jy, or 20.8~mag in the AB system).
The resulting SEDs are those of a representative galaxy with the median F770W flux.
The stellar mass and the SFR scale linearly with this choice, while the colors and all radial trends do not depend on it.

Each radial bin has ten photometric points.
When the central flux of a band is below five times its noise error, we treat that point as an upper limit and follow the CIGALE convention for non-detections.
All ten bands are detected in all five bins, except F090W in the outermost annulus, which is treated as an upper limit.
A calibration floor of 3\% of the flux is added in quadrature to every error.

The parameter uncertainties are derived by refitting bootstrap realizations.
For each of the 50 shared-index realizations (Section~\ref{sec:jwststack}), we build the ten-band SED in every bin and add a random noise draw from the binned noise covariance.
The 16--84\% spread of the fitted parameters over the realizations gives the parameter error.
This error includes both the image noise and the galaxy-to-galaxy variance.
Because the realizations share the same bootstrap indices in all bands and all bins, the correlations between bands and between radii propagate automatically into the errors of the radial gradients.

\subsection{CIGALE Setup} \label{sec:cigale}

We fit the SEDs with CIGALE \citep{Boquien2019} at the fixed median redshift of $z=2.18$.
The model consists of a delayed star formation history, the \citet{BruzualCharlot2003} stellar population models with the \citet{Chabrier2003} IMF and solar metallicity, nebular emission, the two-component attenuation model of \citet[][CF00]{CharlotFall2000}, and the \citet{DraineLi2007} dust emission models \citep{Draine2014}.
Table~\ref{tab:cigale} lists all the model components and the parameter grids.
In the CF00 model, stars younger than 10~Myr are attenuated by both the birth-cloud (BC) and the interstellar medium (ISM) power-law components, while older stars are attenuated only by the ISM component.
We vary the ISM attenuation and fix the ratio $\mu \equiv A_V^{\mathrm{ISM}}/(A_V^{\mathrm{BC}}+A_V^{\mathrm{ISM}})$ to the standard value of 0.44. The results do not depend on this choice.
We measure $A_V$ as the effective (luminosity-weighted) attenuation of the stellar continuum in the $V$ band.
We take the SFR as the instantaneous value of the fitted star formation history.
The dust emission component is required to model the 870~$\mu$m point, and the energy balance in CIGALE ties the absorbed stellar light to the infrared emission.
The ALMA band is registered as a top-hat filter covering the observed frequency range of 335.5--351.5~GHz.
We refined the model grid until the grid step of each parameter was several times smaller than the parameter spread, so that the errors are not limited by the grid resolution.

We also performed the complete analysis with the \citet{Calzetti2000} starburst attenuation curve, keeping everything else identical.
Unlike the CF00 model, this curve attenuates stars of all ages in the same way.
We adopt the CF00 model as the fiducial one because it fits the observed SEDs better in every radial bin, including the central F770W excess left by the Calzetti fits (described below).
Adopting the Calzetti curve instead lowers the stellar masses by 0.16--0.33~dex, the SFRs by a factor of 1.6--4, and the central attenuation by 1.2~mag.
The shapes of the radial profiles are insensitive to this choice (Section~\ref{sec:sedresults}).

\begin{table}
\centering
\footnotesize
\setlength{\tabcolsep}{5pt}
\caption{CIGALE model components and parameter grids}
\label{tab:cigale}
\begin{tabular}{ll}
\hline\hline
Parameter & Values \\
\hline
\multicolumn{2}{c}{Delayed star formation history} \\
\hline
$\tau$ & 0.2, 0.5, 1, 2, 4, 8 Gyr \\
age & 250--3000 Myr, step 125 Myr \\
\hline
\multicolumn{2}{c}{Stellar populations \citep{BruzualCharlot2003}} \\
\hline
IMF & \citet{Chabrier2003} \\
$Z$ & 0.02 \\[2pt]
\hline
\multicolumn{2}{c}{Nebular emission} \\
\hline
$\log U$ & $-2.0$ \\
$Z_{\rm gas}$ & 0.02 \\[2pt]
\hline
\multicolumn{2}{c}{Attenuation \citep{CharlotFall2000}} \\
\hline
$A_V^{\rm ISM}$ & 0--4 mag, step 0.025 mag \\
$\mu$ & 0.44 \\
power-law slope (ISM) & $-0.7$ \\
power-law slope (BC) & $-1.3$ \\[2pt]
\hline
\multicolumn{2}{c}{Attenuation \citep{Calzetti2000}} \\
\hline
$E(B-V)_{\rm lines}$ & 0--1.5, step 0.0125 \\
reduction factor & 0.44 \\
\hline
\multicolumn{2}{c}{Dust emission \citep{DraineLi2007, Draine2014}} \\
\hline
$q_{\rm PAH}$ & 2.5\% \\
$U_{\rm min}$ & 1, 5, 10, 20, 30 \\
$\alpha$ & 2.0 \\
$\gamma$ & 0.1 \\
\hline
\end{tabular}
\tablecomments{All fits are at the fixed redshift $z=2.18$.}
\end{table}

\begin{figure*}
\centering
\figfile[scale=1.0]{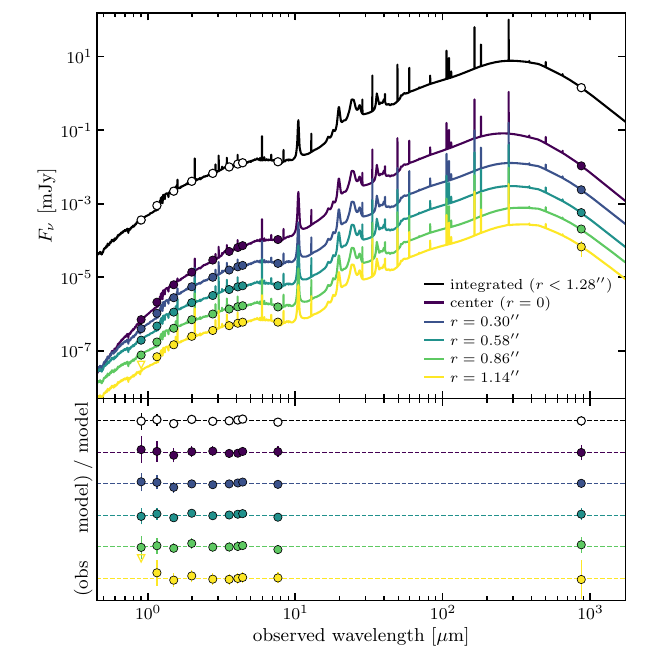}\hfill
\figfile[scale=1.0]{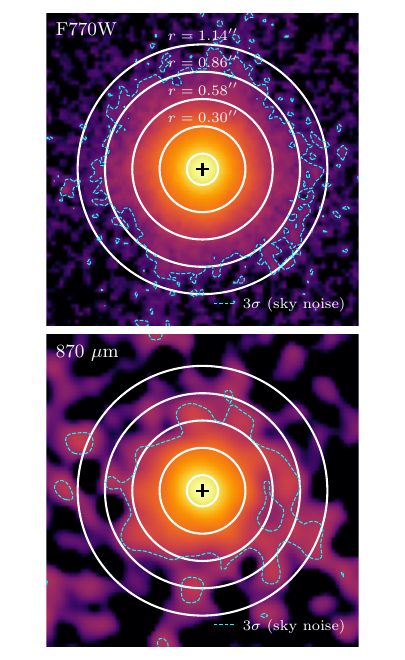}
\caption{
Left panel. Best-fit SEDs of the five radial bins (curves) and the observed photometry (filled circles).
The SEDs are those of a representative galaxy with the median F770W flux of the sample, and the fluxes decrease from the central resolution element ($r=0$, dark purple) to the outermost annulus ($r=1\farcs14$, yellow).
The black curve and the open circles show a separate fit of the integrated photometry over the area covered by the five regions ($r<1\farcs28$).
The radial-bin SEDs are the mean flux densities per $0\farcs04$ pixel, while the integrated SED is the total flux density within the fitted area.
The open downward triangle marks the 5$\sigma$ upper limit (F090W in the outermost annulus).
The bottom sub-panel shows the fractional residuals, $(\mathrm{observed}-\mathrm{model})/\mathrm{model}$, of the integrated SED (top row) and of the five bins, each shifted vertically with the dashed line marking its zero point, from the central resolution element down to the outermost annulus (bottom).
Right panels. The five regions on the F770W (top) and 870~$\mu$m (bottom) stack images.
The cross marks the central resolution element ($r=0$), and the circles show the boundaries of the four annuli.
The labels give the central radius of each annulus ($r=0\farcs30$, $0\farcs58$, $0\farcs86$, and $1\farcs14$), which is used as the radial coordinate of the corresponding SED.
The dashed contour shows the 3$\sigma$ level of the sky noise.
}
\label{fig:sedfits}
\end{figure*}

The left panel of Figure~\ref{fig:sedfits} shows the best-fit SEDs of the five bins together with the observed photometry and the residuals, and the right panels show the five regions on the F770W and 870~$\mu$m stack images.
The models reproduce the ten-band SEDs well at all radii.
With the fiducial CF00 model, the observed and model fluxes agree within 1$\sigma$ in every band and every bin.
With the \citet{Calzetti2000} curve, in contrast, the observed F770W flux in the central resolution element exceeds the best-fit model by 26\% (1.6$\sigma$), and the excess does not disappear when we replace the \citet{BruzualCharlot2003} models with the \citet{Maraston2005} models (2.1$\sigma$), which include a stronger contribution of thermally pulsing AGB stars in the rest-frame near-infrared.
The CF00 model absorbs this excess (the residual becomes 1.5\%, or 0.1$\sigma$) because its power-law ISM curve declines more slowly with wavelength and still attenuates the rest-frame near-infrared, where the Calzetti curve is nearly transparent, so that the red central SED is reproduced by a strongly attenuated old stellar population.
Warm dust heated by an AGN could contribute at rest-frame 2.4~$\mu$m, as suggested for two galaxies of the sample with a red point-like nucleus (Section~\ref{sec:sizecomp}).
However, because the excess appears only with the Calzetti curve and F770W is the only band in this wavelength range, we do not include an AGN component in the fit.
We quantify the possible AGN contribution and its impact on our results in Section~\ref{sec:agntest}.

As a check of the radial binning, we also fitted the integrated photometry.
We summed the stacked images over the exact area covered by the five regions and fitted this ten-band SED separately with the same setup.
The integrated photometry and its best-fit model are shown in Figure~\ref{fig:sedfits}.
The stellar mass and the SFR summed over the five bins agree with the values of the integrated fit to within 5\% and 1\%.
The effective $A_V$ of the integrated fit is 2.5~mag, which lies between the central and outer values of the resolved profiles.
Between F770W and 870~$\mu$m the models are not constrained by data, but the sum of the five bin models and the integrated model still agree to within 10\% in this range.
With the Calzetti curve the difference in this range is larger, with a median of 60\%, although the summed masses and SFRs agree in the same way.
The radial-bin fits and the integrated fit therefore give a consistent picture.

\subsection{Radial Profiles of Stellar Mass, SFR, and Attenuation}
\label{sec:sedresults}

\begin{figure*}
\centering
\figfile[scale=1.0]{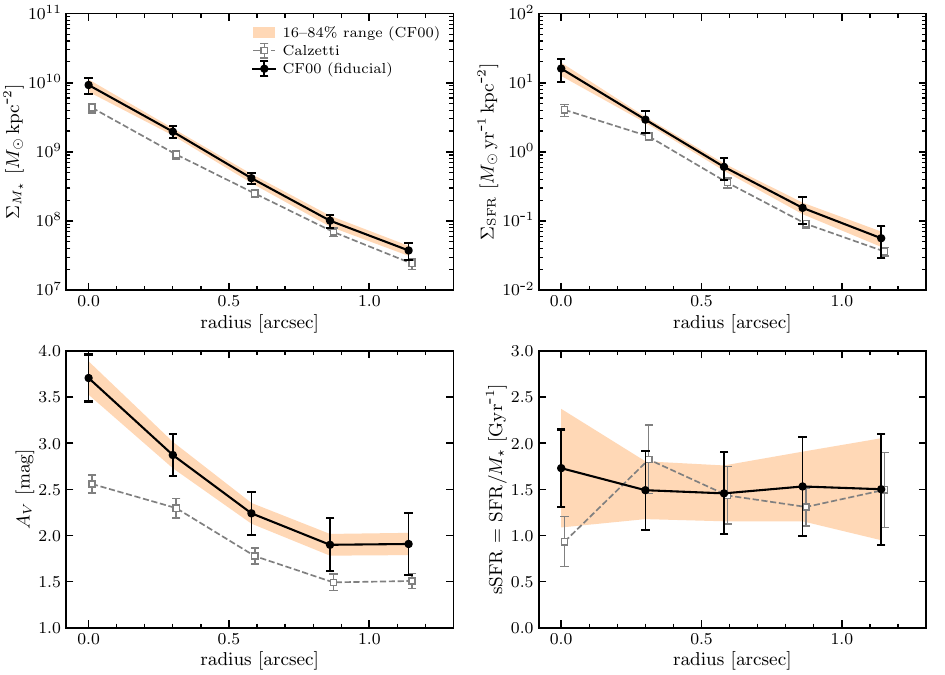}
\caption{
Radial profiles of the stellar mass surface density (top left), the SFR surface density (top right), the attenuation $A_V$ (bottom left), and the sSFR (bottom right) from the CIGALE fits to the radial-bin SEDs.
Black circles show the fiducial CF00 fits and gray open squares the fits with the \citet{Calzetti2000} curve, which are shifted slightly in radius for clarity.
The error bars on the central values show the Bayesian uncertainties of the fits, and the shaded bands show the 16--84\% spread of the 50 bootstrap realizations, which we adopt as the parameter errors.
The stellar mass and SFR surface densities are those of a representative galaxy with the median F770W flux of the sample.
}
\label{fig:sedprofiles}
\end{figure*}

Figure~\ref{fig:sedprofiles} shows the radial profiles of the stellar mass surface density $\Sigma_{M_\star}$, the SFR surface density $\Sigma_\mathrm{SFR}$, the attenuation $A_V$, and the specific SFR ($\mathrm{sSFR}=\mathrm{SFR}/M_\star$).
The $\Sigma_{M_\star}$ and $\Sigma_\mathrm{SFR}$ profiles both decline steeply, by a factor of about 250 and 280, respectively, from the center to the outermost annulus ($r=9.4$~kpc).
The central resolution element reaches $\Sigma_{M_\star}\simeq9.3\times10^{9}~\Msun~\mathrm{kpc^{-2}}$ and $\Sigma_\mathrm{SFR}\simeq16~\Msun~\mathrm{yr^{-1}~kpc^{-2}}$.
These central values are still averaged over the PSF and are therefore lower limits on the intrinsic central surface densities.
When the five regions are placed in the plane of $\Sigma_\mathrm{SFR}$ against $\Sigma_{M_\star}$, they span about two orders of magnitude in both quantities and follow the resolved main sequence of star formation found for star-forming galaxies at $0.7<z<1.5$ \citep{Wuyts2013}.
Each region thus forms stars at the normal rate for its local stellar mass surface density.

The attenuation is highest at the center.
$A_V$ decreases from $3.71\pm0.18$~mag at the center to $1.91\pm0.12$~mag at $r\gtrsim7$~kpc, with a slope of $-0.195\pm0.021~\mathrm{mag~kpc^{-1}}$.
In contrast, the sSFR profile is flat.
The values stay between 1.5 and 1.7~$\mathrm{Gyr^{-1}}$ at all radii, and the slope of $\log(\mathrm{sSFR})$ with radius is $-0.005\pm0.016~\mathrm{dex~kpc^{-1}}$, which is consistent with zero.
The central resolution element has the highest value ($1.73\pm0.64~\mathrm{Gyr^{-1}}$), but its uncertainty is large and the profile is consistent with a constant sSFR.
With the Calzetti curve the central bin is instead the lowest of the five ($0.93\pm0.27~\mathrm{Gyr^{-1}}$), so the central sSFR is the quantity most sensitive to the attenuation law.
In neither case is the deviation from a flat profile significant.
The ongoing star formation therefore follows the existing stellar mass distribution at all radii, from the compact center to the outer disk.

Four caveats apply to these profiles.
First, the stack mixes galaxies at different redshifts and is fit at the median redshift, so each filter samples a slightly different rest-frame wavelength for each galaxy and the composite SED is smoothed by this mixing.
Second, stacking in angular units mixes galaxies with different sizes, which dilutes real gradients.
The measured gradients are therefore lower limits on the typical gradients of the individual galaxies.
Third, the absolute values of $\Sigma_{M_\star}$ and $\Sigma_\mathrm{SFR}$ scale with the adopted median F770W flux, while $A_V$, the sSFR, and all radial trends do not depend on this choice.
Fourth, the derived $A_V$ is an effective attenuation of the light that escapes the galaxy, and stars behind optically thick dust do not contribute to the observed SEDs even in F770W.
The stellar mass surface densities should therefore be taken as lower limits.
We discuss this point in Section~\ref{sec:disc_sizes}.
\subsection{Impact of an AGN Component} \label{sec:agntest}

With the fiducial CF00 model, the ten-band SEDs are reproduced without an AGN component, and there is no sign that one is needed.
The central F770W excess appears only with the Calzetti curve, and warm dust around an obscured AGN is an alternative way to explain it.
Because the CF00 model is not necessarily the correct description of the attenuation, we test this alternative on the Calzetti fits.
We repeat the ten-band fits with the Calzetti curve, adding the SKIRTOR clumpy torus models \citep{Stalevski2012, Stalevski2016} implemented in CIGALE.
All the other modules and grids are kept identical to the Calzetti setup, and the only new free parameter is the AGN fraction $f_\mathrm{AGN}$ of the total infrared luminosity, with a grid of eight values from 0 to 0.7.
The torus geometry parameters are fixed at typical values, and we run two fits with different viewing angles, a face-on view ($i=30\arcdeg$, type 1), in which the accretion disk is seen directly, and an edge-on view ($i=70\arcdeg$, type 2), in which the disk is hidden by the torus.

The two geometries give opposite results.
In the face-on case, the best-fit $f_\mathrm{AGN}$ is zero and the $\chi^2$ of the central bin does not change, because the unobscured disk continuum would add flux to the shorter-wavelength bands that the stellar models already reproduce well.
In the edge-on case, the torus emission peaks in the mid-infrared and can be added without disturbing the other bands.
The best-fit $f_\mathrm{AGN}$ of the central bin is 0.3, the F770W excess decreases from 26\% to 5\%, and the $\chi^2$ improves from 2.4 to 0.6.
Even in this case, however, the marginalized estimate is $f_\mathrm{AGN}=0.20\pm0.16$, which differs from zero by only 1.2$\sigma$.
An obscured AGN can therefore absorb the central F770W excess, but the data do not require it.
With F770W as the only band in this wavelength range, the current photometry cannot tell whether the excess seen with the Calzetti curve comes from the age-dependent attenuation, from an obscured AGN, or from a statistical fluctuation.

This choice does not affect our results.
In both geometries, the changes of $M_\star$, SFR, and $A_V$ are a few percent in the central resolution element and stay well within the parameter errors in all bins.
The radial trends presented in Section~\ref{sec:sedresults}, namely the central concentration of the stellar mass and SFR and the flat sSFR profile, are therefore insensitive to the AGN component.

\section{Discussion} \label{sec:discussion}

\subsection{The Stellar Mass Distribution Is More Compact than the
Near-infrared Light Suggests} \label{sec:disc_sizes}

JWST and ALMA studies of dusty star-forming galaxies have compared the stellar light at 4.4~$\mu$m with the dust continuum at 870~$\mu$m \citep[][see also \citealt{Bodansky2026} for the 1.1~mm continuum]{Chen2022,Gillman2024,Hodge2025}.
The dust emission remains more compact than the rest-frame near-infrared light observed with NIRCam.
For example, \citet{Hodge2025} found that the F444W sizes of submillimeter galaxies at $z\sim3$ are $78\pm21$\% larger than their 870~$\mu$m sizes.
Our sample shows the same behavior.
The F444W sizes exceed the 870~$\mu$m sizes by about 30\% (Section~\ref{sec:sizecomp}).
One could interpret this as a compact starburst inside a larger stellar body.

Our F770W measurements with MIRI do not support this interpretation.
The F770W profile matches the 870~$\mu$m profile, and the nine-band profiles become more compact continuously from F090W through F444W to F770W (Figure~\ref{fig:profiles}).
The resolved SEDs explain why.
The attenuation rises to $A_V=3.7$~mag in the center, which dims the rest-frame 1.4~$\mu$m light by $\simeq$1.9~mag and the rest-frame 2.4~$\mu$m light by $\simeq$1.3~mag, so the F444W light loses 0.6~mag more than the F770W light.
This central value is an average over the central resolution element ($\mathrm{FWHM}=0\farcs28$, or 2.3~kpc), so the attenuation at the very center is likely even higher.
In dusty galaxies like our targets, even F444W is affected by the central dust attenuation and does not trace the stellar mass distribution \citep[see also][]{Kamieneski2023}.
The apparent difference between the 4.4~$\mu$m and 870~$\mu$m sizes is thus mostly another sign of the central dust concentration, not a true offset between the star formation and the stellar mass distributions.
The same picture is found in submillimeter galaxies at higher redshift.
\citet{Li2026} studied 12 galaxies at $z\sim3$ with NIRCam, MIRI, and ALMA data.
Their sample is selected at 870~$\mu$m.
In their sample, the strong obscuration of the center makes the stellar size appear larger at rest-frame optical wavelengths, and this effect becomes small in F770W.
Their stellar sizes, corrected for the attenuation, agree with their 870~$\mu$m sizes with a ratio of $1.0\pm0.4$, close to our value of 0.95.
They concluded that rest-frame wavelengths longer than $\simeq$1.6~$\mu$m are needed to recover the stellar morphology, which is what F770W provides at $z\sim2$.
The two samples are complementary.
The stellar masses are similar, but the median 870~$\mu$m flux density of their sample is about three times higher than that of ours, and their attenuations are also higher.
Our sample is instead selected by stellar mass and represents the general population of massive star-forming galaxies (Section~\ref{sec:sample}).
The agreement between the stellar mass and the dust distributions therefore holds both for the brightest 870~$\mu$m sources and for the general population.

The $A_V$ values derived from the SED fits deserve a note.
The dust mass surface densities of similar galaxies imply an extremely large attenuation, with $A_V$ of several hundred magnitudes through the central few kiloparsecs \citep{Simpson2017}.
This column value and our effective value of $A_V=3.7$~mag do not conflict with each other, because they measure different things.
The effective value describes the light that escapes from the galaxy, and the two values can differ by orders of magnitude if the dust is patchy, with optically thick clumps and diffuse dust.
In fact, the F090W flux is detected even in the central resolution element, so some sightlines toward the center must be almost free of dust.
Stars behind the optically thick clumps, however, do not contribute to the observed SEDs at any wavelength up to F770W.
The stellar masses derived with CIGALE are therefore lower limits, and part of the stellar mass can still be hidden even in the rest-frame $K$ band.
The central surface density, however, leaves little room for a large hidden mass, because the observed value is already similar to the central densities of the likely descendants (Section~\ref{sec:disc_growth}).
\subsection{Bulge Growth in the Final Phase} \label{sec:disc_growth}

The stacked 870~$\mu$m and F770W profiles agree out to $\simeq$8~kpc (Section~\ref{sec:profiles}), and the sSFR profile is flat at about 1.5~$\mathrm{Gyr^{-1}}$ (Section~\ref{sec:sedresults}).
The obscured star formation is therefore distributed like the existing stellar mass, and every radius grows at nearly the same relative rate.
In other words, these galaxies are growing without changing their structure.

This behavior is not what is expected during a compaction event, in which star formation concentrates in the center and the bulge fraction rises quickly.
Instead, the bulge is already in place.
The decomposition of the F770W stack gives $B/T=0.50$ with a bulge effective radius of 0.8~kpc (Section~\ref{sec:bulgedisk}), and this bulge size agrees with the bulges found in submillimeter galaxies at F444W \citep[median 0.7~kpc,][]{Chen2022}.
A high bulge fraction in the rest-frame near-infrared is also found in the general population of massive star-forming galaxies at this epoch \citep{Benton2024}.
\citet{Lang2014} decomposed the stellar mass maps of star-forming galaxies at $1.5<z<2.5$, which were reconstructed from resolved stellar population modeling, and found that the bulge fraction reaches $B/T=0.4$--0.5 above $\log(M_\star/\Msun)=11$.
We note that our $B/T$ is a light-weighted value at rest-frame 2.4~$\mu$m, and a difference in the mass-to-light ratio between the bulge and the disk would shift the mass-weighted value.
The agreement with the mass-weighted measurement of \citet{Lang2014} suggests that this shift is not large.

The flat sSFR sets the growth timescale.
The mass doubling time is $1/\mathrm{sSFR}=0.6$--0.7~Gyr at all radii.
The stellar surface density of the central resolution element is already $\Sigma_{M_\star}\simeq9.3\times10^{9}~\Msun~\mathrm{kpc^{-2}}$, which is similar to the central densities of quiescent galaxies at the same epoch \citep{vanDokkum2014,Barro2017}.
The dense center of a quiescent galaxy is thus already in place.
If the current star formation continued for much longer than one doubling time, the centers would become denser than those of the quiescent population at $z\simeq1.5$--2.
The star formation in the central region is therefore likely to stop within about 1~Gyr, and the galaxies are natural progenitors of that population.

The molecular gas adds an interesting contrast.
\citet{Tadaki2023} compared the CO $J=3$--2 line, the 870~$\mu$m continuum, and the F444W light in massive star-forming galaxies at $z=2.2$--2.5.
The 870~$\mu$m emission is about 40\% more compact than the CO emission, while the CO emission is as extended as the rest-frame 1.4~$\mu$m light.
One interpretation of this contrast is that the obscured star formation is more concentrated than the molecular gas that feeds it, so that the gas depletion time is shorter in the center than in the outer disk.
In this interpretation, the two regions may correspond to the two modes of star formation seen in the relation between the gas and SFR surface densities \citep{Daddi2010,Genzel2010}.
The center forms stars in a starburst-like mode, while the outer disk forms stars in the normal disk mode.
There is, however, another interpretation of the size difference between the CO and the 870~$\mu$m emission.
A temperature gradient of the dust boosts the central surface brightness of the continuum, and radiative transfer models show that temperature and optical depth gradients can produce the apparent size difference between the dust continuum and the CO emission even when the dust and the gas are co-spatial \citep{CalistroRivera2018,Boogaard2026}.
In that case the size difference does not imply a shorter depletion time in the center.

Our results add one more step to this comparison.
Because the F770W and 870~$\mu$m emissions share the same spatial distribution, the F770W light is also likely more compact than the CO emission, although the two have not yet been compared within the same sample.
This means that the ratio of the gas mass to the stellar mass is lower in the central region than in the outer disk.
The center has already converted most of its gas into stars, which fits the picture that the bulge is close to the end of its growth, while the outer disk still keeps a larger gas reserve.
A radial gradient of the CO excitation could affect this comparison, but the correction goes in one direction.
The CO $J=3$--2 excitation is likely higher in the center, so the true gas distribution would be even more extended than the observed CO emission, which strengthens this conclusion.

\subsection{Different Phases of Bulge Formation in the Sample}
\label{sec:disc_phases}

The flat sSFR profile also tells us when the bulge was built.
The current star formation preserves the shape of the galaxies, so the bulge fraction of $B/T=0.50$ cannot be raised in the present phase.
The bulge must have been assembled when the galaxies were less massive.
Less massive star-forming galaxies, however, are disks \citep{Wuyts2011}.
In the rest-frame near-infrared, their median S\'ersic index stays at $n\simeq1.4$ up to $\log(M_\star/\Msun)\simeq10.5$ and rises to $n\simeq2$--3 above $\log(M_\star/\Msun)=11$ \citep{Martorano2025}.
Our median index of $n=2.4$ (Section~\ref{sec:galfit}) matches the massive end of this relation.
If our galaxies grew from such disk-dominated progenitors, the bulge must have been built while the galaxies grew from $\log(M_\star/\Msun)\simeq10.5$ to 11.
In that mass range, the star formation must have been much more concentrated than the existing disk.
A simple estimate shows that roughly three quarters of the stars formed in this mass interval must form in the central component to raise $B/T$ from $\simeq$0.1 to 0.5.

There are other ways to raise the central concentration.
Mergers can drive gas to the center and build a bulge \citep{Hopkins2008}.
\citet{LillyCarollo2016} proposed that no special event is needed at all.
In their picture, star-forming disks grow from the inside out, the disks were smaller at earlier times, and the old inner disk naturally remains as a dense core.
Two observations favor a phase of concentrated star formation in our case.
First, inside-out growth predicts an sSFR profile that rises outward, while the observed profile is flat.
Second, the required mode of star formation, much more compact than the existing stellar body, is directly seen in two galaxies of our sample, U4-34617 and U4-36247.
The two pictures are not exclusive.
The smooth growth of \citet{LillyCarollo2016} may describe most of the lifetime of these galaxies, with short compaction episodes on top of it.

These two galaxies belong to the four outliers in Section~\ref{sec:sizecomp}, which suggest that the galaxies are not all in the same phase.
With their compact dust emission and extended stellar light, they look like systems in the middle of a compaction event, in which gas falls to the center and feeds a compact starburst \citep[e.g.,][]{DekelBurkert2014,Zolotov2015}.
They are also the two least massive galaxies of the sample ($\log(M_\star/\Msun)=11.0$), just above the mass range in which the bulge should be built.
U4-34617, where half of the 870~$\mu$m flux is still in an extended component, may be at an earlier stage than U4-36247, where almost all of the dust emission is already compact.
The majority of the sample, in which the star formation follows the stellar mass, would then be in the phase after compaction, when the bulge is largely assembled.
Finally, the two galaxies with a red point-like nucleus (U4-42571 and U4-36685) may host an AGN.
In the compaction picture, the central gas concentration also feeds the black hole, and AGN activity is expected toward the end of the compact star-forming phase.

\section{Summary} \label{sec:summary}

We have studied bulge formation in 16 massive dusty star-forming galaxies at $z\sim2$.
We combined JWST MIRI F770W imaging, JWST NIRCam imaging in eight bands, and ALMA 870~$\mu$m data at a matched angular resolution.
Our main results are as follows.

\begin{enumerate}
\item We measured the sizes of the 16 galaxies with GALFIT.
The circularized effective radii at rest-frame 2.4~$\mu$m agree with those of the 870~$\mu$m dust emission, with a median ratio of 0.95, and the centroids of the two emissions coincide, with a median separation of $0\farcs04$ (0.3~kpc).
The F770W sizes are typically 23\% smaller than the F444W sizes.
The stellar mass distribution is thus as compact as the dusty star-forming region.
\item Four galaxies deviate from this overall agreement.
In U4-34617 and U4-36247, the dust emission is 3 to 6 times more compact than the rest-frame 2.4~$\mu$m light, as expected during a wet compaction event.
U4-42571 and U4-36685 host a red point-like nucleus, which suggests warm dust heated by an AGN.
\item We stacked the 16 galaxies in the ten bands, matched to the F770W stack PSF ($\mathrm{FWHM}=0\farcs28$).
The stacked 870~$\mu$m and F770W profiles agree closely out to $\simeq$8~kpc.
The profiles of the nine JWST bands become more compact toward longer wavelengths, which traces a continuous radial color gradient, and even F444W is affected by the dust attenuation in the central region.
\item A bulge plus disk decomposition of the stacked F770W profile gives a bulge-to-total ratio of $B/T=0.50$ with a bulge effective radius of 0.8~kpc and a disk effective radius of 2.4~kpc.
The bulge is already in place in these galaxies.
\item We fit the SEDs in five radial bins with CIGALE using the ten bands.
The five radial bins follow the resolved main sequence between the stellar mass and SFR surface densities.
The attenuation decreases from $A_V=3.7$~mag at the center to 1.9~mag at $r\gtrsim7$~kpc.
The sSFR profile is flat at about 1.5~$\mathrm{Gyr^{-1}}$, which means that the obscured star formation follows the existing stellar mass distribution, so the galaxies grow without changing their structure.
The mass doubling time is 0.6--0.7~Gyr, and the central stellar surface density is already similar to that of quiescent galaxies at the same epoch.
These galaxies are in the final phase of bulge formation and are natural progenitors of compact quiescent galaxies at $z\simeq1.5$--2.
\end{enumerate}

Our sample may thus catch massive galaxies in several phases of bulge formation.
Most of them are in the phase where the bulge is in place and the remaining star formation builds the whole system at the same relative rate.
U4-34617 and U4-36247 appear to be in the middle of the bulge-building compaction, and the AGN activity suggested in U4-42571 and U4-36685 may mark the end of that phase.
This picture can be tested with statistics.
In a mass-limited sample with both MIRI and ALMA 870~$\mu$m imaging, the fraction of galaxies whose dust emission is much more compact than the rest-frame $K$-band light gives the fraction of galaxies in the compaction phase.
This fraction translates into the time spent in that phase.
Such a survey will tell how long the compact star formation lasts and at which stellar mass it happens most often.

\begin{acknowledgments}
The author thanks the anonymous referee for the careful reading of the manuscript and for the constructive comments that improved this paper.
The author thanks Ian Smail for constructive comments on the dust geometry and the interpretation of the compact component.
This paper makes use of the following ALMA data: ADS/JAO.ALMA\#2017.1.01027.S and \#2012.1.00245.S.
ALMA is a partnership of ESO (representing its member states), NSF (USA), and NINS (Japan), together with NRC (Canada), NSTC and ASIAA (Taiwan), and KASI (Republic of Korea), in cooperation with the Republic of Chile.
The Joint ALMA Observatory is operated by ESO, AUI/NRAO, and NAOJ.
This work is based on observations made with the NASA/ESA/CSA James Webb Space Telescope.
All the JWST data used in this paper can be found in MAST: \dataset[doi:10.17909/qqqh-kv91]{https://doi.org/10.17909/qqqh-kv91}.
The data products presented herein were retrieved from the Dawn JWST Archive (DJA). DJA is an initiative of the Cosmic Dawn Center (DAWN), which is funded by the Danish National Research Foundation under grant DNRF140.
The author used Claude (Anthropic) for assistance with language editing and code refinement. The author takes full responsibility for the content of this manuscript.
K.T. acknowledges support from JSPS KAKENHI Grant Number JP 23K03466.
K.T. was supported by the ALMA Japan Research Grant of NAOJ ALMA Project, NAOJ-ALMA-403.
\end{acknowledgments}

\facilities{ALMA, JWST(NIRCam, MIRI)}

\software{CASA \citep{CASA2022}, UVMULTIFIT \citep{MartiVidal2014},
GALFIT \citep{Peng2002,Peng2010},
STPSF \citep{Perrin2014}, photutils \citep{Bradley2026},
CIGALE \citep{Boquien2019}, astropy \citep{Astropy2022}}

\appendix

\section{Calibration of the F770W Size Measurements} \label{app:calib}

We test how well the GALFIT analysis of Section~\ref{sec:sizes} recovers compact sizes with injection and recovery simulations, and we use them to estimate the systematic uncertainty tied to the PSF model.
The 21 PSF stars are split into two independent halves, with 11 and 10 stars, and a hybrid PSF is built from each half in the same way as in Section~\ref{sec:psf}.
Simulated galaxies are created with one PSF and fit with the other.
The small difference between the two PSFs then propagates a realistic PSF uncertainty into the recovered parameters.

We run the simulations for the three galaxies with the smallest measured F770W sizes (U4-190, U4-36685, and U4-42571), for which the PSF model matters most.
For each galaxy, we fix the magnitude, the axis ratio, and the position angle to the best-fit values, and we vary the input effective radius over a grid from $0\farcs008$ to $0\farcs12$ (0.07 to 1.0~kpc).
Each model is injected into source-free sky cutouts of the same mosaic with a matching exposure time, using 50 positions for U4-190 and U4-36685 and 5 positions for U4-42571, and is refit with $n=1$ in the same way as the real data.

The recovered axis ratios separate into two branches.
Fits on the good branch ($q_\mathrm{out}>0.5$) recover the input parameters, while fits on the catastrophic branch ($q_\mathrm{out}<0.2$) collapse to very elongated shapes and wrong sizes.
The catastrophic branch appears only for very compact inputs.
Inputs with $R_\mathrm{e}\lesssim0\farcs08$ fail in 50 to 90\% of the trials, and the failure rate drops to zero at $R_\mathrm{e}\gtrsim0\farcs1$.
All the real fits of the 16 galaxies lie on the good branch, so we derive the calibration from the good branch only.

\begin{figure*}
\centering
\figfile[scale=1.0]{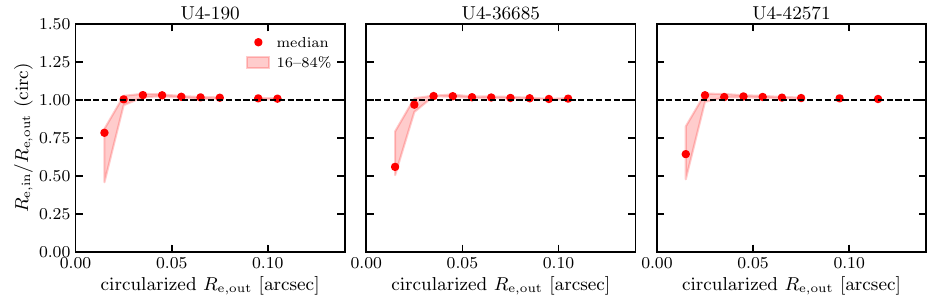}
\caption{
Injection-and-recovery calibration of circularized effective radii, $R_\mathrm{e}\sqrt{q}$, for the three most compact galaxies in the sample (from left to right, U4-190, U4-36685, and U4-42571), under the PSF-mismatch procedure described in the text.
For each target, the red circles and shaded band show the median and 16--84th percentile range in bins of $0\farcs01$, and the dashed line marks unity.
}
\label{fig:calib}
\end{figure*}

Figure~\ref{fig:calib} shows the result for all three galaxies.
We measure the ratio of the input to the recovered circularized radius as a function of the recovered value, because the recovered value is the quantity available for the real data.
For recovered radii above $0\farcs035$ (0.3~kpc), the ratio stays between 1.01 and 1.03 for all three galaxies.
At the measured sizes of the three galaxies themselves (0.53 to 0.67~kpc), the ratio is 1.01 to 1.02 with a 16--84\% spread below 1\%.
The measured sizes are therefore underestimated by only 1 to 2\% under a realistic PSF uncertainty.
This systematic is comparable to the statistical uncertainties in Table~\ref{tab:sample}, and we do not apply any correction to the measured sizes.
Recovered radii below $0\farcs024$ (0.2~kpc) can no longer be mapped back to a unique input size, but no galaxy in our sample falls in this range.

\bibliographystyle{apj}
\bibliography{ref}

\end{document}